\documentstyle[12pt]{article}

\begin{document}

\baselineskip=6mm

\newcommand{\TeV}{{\rm TeV}}
\newcommand{\GeV}{{\rm GeV}}
\newcommand{\MeV}{{\rm MeV}}
\newcommand{\keV}{{\rm keV}}
\newcommand{\eV}{{\rm eV}}

\newcommand{\be}{\begin{equation}}
\newcommand{\ee}{\end{equation}}
\newcommand{\bea}{\begin{eqnarray}}
\newcommand{\eea}{\end{eqnarray}}
\newcommand{\ba}{\begin{array}}
\newcommand{\ea}{\end{array}}
\newcommand{\bmat}{\left(\ba}
\newcommand{\emat}{\ea\right)}
\newcommand{\refs}[1]{(\ref{#1})}
\newcommand{\ler}{\stackrel{\scriptstyle <}{\scriptstyle\sim}}
\newcommand{\ger}{\stackrel{\scriptstyle >}{\scriptstyle\sim}}
\newcommand{\lag}{\langle}
\newcommand{\rag}{\rangle}
\newcommand{\ns}{\normalsize}
\newcommand{\cm}{{\cal M}}
\newcommand{\gr}{m_{3/2}}
\newcommand{\db}{\bar{\partial}}
\newcommand{\fb}{\bar{f}}
\newcommand{\wb}{\bar{w}}
\newcommand{\Mb}{\bar{M}}
\newcommand{\p}{\partial}
\newcommand{\nnu}{\nonumber}
\def\a{\alpha}
\def\b{\beta}
\def\g{\gamma}
\def\c{\chi}
\def\d{\delta}
\def\e{\epsilon}
\def\l{\lambda}
\def\x{\xi}
\def\r{\rho}
\def\s{\sigma}
\renewcommand{\Huge}{\Large}
\renewcommand{\LARGE}{\Large}
\renewcommand{\Large}{\large}

\begin{titlepage}
\title{\bf Axino Mass in Supergravity Models\\
                          \vspace{-4cm}
                          \hfill{\ns TUM-HEP 215/95\\}
                          \hfill{\ns IC/95/29\\}
                          \hfill{\ns hep-ph/9503233\\[.5cm]}
                          \hfill{\ns February 1995}
                          \vspace{2cm} }

\author{E.~J.~Chun\thanks{Email:chun@ictp.trieste.it}\\[0.2cm]
        {\ns International Center for Theoretical Physics}\\
        {\ns P.~O.~Box 586, 34100 Trieste, Italy}\\[2mm]
        {\ns and}\\[2mm]
        A.~Lukas\thanks{Email:alukas@physik.tu-muenchen.de}\\[0.2cm]
        {\ns Physik Department}\\
        {\ns Technische Universit\"at M\"unchen}\\
        {\ns D-85747 Garching, Germany}\\
        {\ns and}\\[0.1cm]
        {\ns Max--Planck--Institut f\"ur Physik}\\
        {\ns Werner--Heisenberg--Institut}\\
        {\ns P.~O.~Box 40 12 12, Munich, Germany}}

\date{}
\maketitle

\begin{abstract} \baselineskip=6mm
We analyze the mass of the axino, the fermionic superpartner of the axion,
in general supergravity models incorporating a Peccei--Quinn--symmetry
and determine the cosmological constraints on this mass. In particular,
we derive a simple criterion to identify models with an LSP--axino
which has a mass of $O(m_{3/2}^2/f_{PQ})=O($keV) and can serve as a
candidate for (warm) dark matter. We point out that such models have
very special properties and in addition, the small axino mass has to
be protected against radiative corrections by demanding small couplings
in the Peccei--Quinn--sector. Generically, we find an axino mass of order
$m_{3/2}$. Such masses are constrained by the requirement of an axino
decay which occurs before the decoupling of the ordinary LSP. Especially,
for a large Peccei--Quinn--scale $f_{PQ}>10^{11}$ GeV this constraint
might be difficult to fulfill.
\end{abstract}

\thispagestyle{empty}
\end{titlepage}

The implications of axions have been examined extensively since their
existence was suggested by an attractive mechanism for resolving the strong
CP problem~\cite{ksvz,dfsz,kim}. Even though axions are very weakly
interacting, their astrophysical and cosmological effects are strong enough
to narrow down the window of the Peccei--Quinn--scale $f_{PQ}$ to
$10^{10}\; \GeV < f_{PQ} < 10^{12}\; \GeV$~\cite{kt}.
On the same footing the axino as the supersymmetric partner of the axion
can play an important role in astrophysics and cosmology~\cite{rtw}.
An interesting feature is that axinos may receive a mass of order keV which
would render them a good candidate for warm dark matter.
If axinos are heavier than a few keV they have to decay fast enough not
to upset any standard prediction of big--bang cosmology.  Given the
weakness of their interactions, a constraint on their lifetime put a rather
severe limit on the lower bound of their mass. Therefore it is very
important to know the axino mass in discussing the cosmological implications
of supersymmetric axion models.

In global supersymmetry (SUSY) the calculation of the axino mass
was performed in refs.~\cite{tw,nie}.  In this paper, we will provide the
computations in models with local supersymmetry (supergravity).
Some partial results have been obtained in refs.~\cite{gy,ckn}.
In the case of spontaneously--broken global SUSY the axino mass is
of the order $m_{3/2}^2/f_{PQ} \sim \keV$ where $m_{3/2} (\ler 1\TeV)$ is
taken to be the global SUSY--breaking scale~\cite{tw,nie}.
On the contrary, in the context of supergravity, the axino mass can be
of order $m_{3/2}$ as first noticed in ref.~\cite{gy}. Soon after this it
was realized that the axino mass is truly model--dependent and the global
SUSY value $m_{3/2}^2/f_{PQ}$ may be obtained in supergravity models as
well~\cite{ckn}. We will extend those results in a generic treatment of
supergravity models also including radiative corrections.

The prime motivation for supergravity is well--known.
Realistic supersymmetric generalizations of the standard model are based on
local SUSY spontaneously broken in a so--called hidden sector at a mass
scale of order $M_S \sim 10^{11} \GeV$~\cite{nil}.  The induced SUSY
breaking scale in the observable sector is determined by
a value of the order of the gravitino mass $m_{3/2} \sim M_S^2/M_{P}$ where
$M_{P}$ is the Planck scale. Axionic extensions of the minimal
supersymmetric standard model (MSSM) inevitably incorporate an extra
sector which provides spontaneous breaking of the Peccei--Quinn
$U(1)$--symmetry at the scale $f_{PQ}$.
This sector (PQ--sector) is considered as a part of the observable sector.
In the framework of effective supergravity theories with a Lagrangian
composed out of a global SUSY part and soft terms the hidden sector
dependences are encoded in the soft terms.  We will rely mostly on this
effective approach as it makes the calculations tractable.

A color anomaly in the PQ--sector can be introduced in two ways. The fields
$S$ in this sector can be coupled to the standard Higgs doublets $H_1$,
$H_2$ of the MSSM like $gSH_1H_2$ (DFSZ--axion)~\cite{dfsz,chun} or to
new heavy quarks $Q_1$, $Q_2$ like $gSQ_1Q_2$ (KSVZ--axion)~\cite{ksvz}.
Our analysis of axino mass will be concerned with the tree level result
in the effective theory which is obtained after breaking the PQ--symmetry
and SUSY and should therefore not depend on which implementation of the
axion is chosen. We will find that this mass crucially depends on the
structure of the PQ--sector as well as on the hidden sector. It should,
however, be mentioned that in addition the above couplings and the
corresponding trilinear soft terms lead to a one--loop radiative mass of
order $m_{\tilde{a},{\rm loop}}\simeq 3g^2Am_{3/2}/16\pi^2$ ($A$ is the
trilinear soft coupling)~\cite{my,gy}. For the DFSZ--axion this
contribution is clearly negligible since the coupling $g$ has to be quite
small ($g\sim 10^{-7}$) in order to generate reasonable Higgs--masses.
In the KSVZ--implementation there is no such restriction on $g$ and the
above one--loop contribution can modify the result for the axino mass to be
obtained below.

\bigskip

An interesting observation is that the axino mass depends on
whether the PQ--sector admits additional accidental zero modes on the
global SUSY level. As a general statement we find that the
axino mass cannot be bigger than $O(\gr )$ if the axion mode is
the only global zero mode in the PQ--sector.
In order to see this, it is useful to analyze the full supergravity
Lagrangian.  Later, we will confirm this result by using the effective
Lagrangian approach.

The PQ--sector consists of an arbitrary number of singlets $S^a$ with
charge $q_a$ under the PQ--symmetry and provides its spontaneous breaking
at the scale $f_{PQ}=(\sum_aq_a^2|v^a|^2)^{1/2}$, where $v^a=\lag s^a \rag$.
The axion multiplet
$\Phi$ is given by $\Phi =\sum_aq_av^aS^a/f_{PQ}$. Its component field
content reads $\Phi\sim (s+ia,\tilde{a})$ with the axion $a$, the saxion
$s$ and the axino $\tilde{a}$.
Supersymmetry is broken mainly by the hidden sector with singlet fields
$Z^i$. To simplify the argument we assume minimal kinetic term for the PQ
fields as well as for the hidden sector fields. The scalar potential reads
\be
 V = M_P^2 \exp[G/M_P^2]
         \left( \sum_i G_i G_i^* + \sum_a G_a G_a^* - 3 M_P^2 \right)\,,
\ee
where $G = K + M_P^2 \ln|W/M_P^3|^2$ and $W = W(S^a) + W(Z^i)$.
Spontaneous supersymmmetry breaking implies $\lag G_i \rag \simeq M_P$
for some $i$ and  generation of a gravitino mass,
$m_{3/2} \simeq M_P \exp[G/2M_P^2]$.  We also have to admit the
possibility that $\lag G_a \rag \ler M_P$ for some $a$.
We now look at  the minimization condition
$\lag V_i \rag = 0$ to estimate the axino mass. Vanishing of the
cosmological constant $\lag V \rag =0$ is assumed. Then we have
\be
  \lag V_a\rag = \lag G_{ab} G_b^* + G_{ai} G_i^* + G_a^* \rag \,.
\ee
If we take a massive mode $S^a$ in the PQ--sector, $\lag G_{ab}\rag$ is
dominated by
\be
  \lag G_{ab} \rag
            \simeq \lag M_P^2 { W_{ab} \over W} \rag
            \simeq {f_{PQ} \over m_{3/2}} \delta_{ab} \,,
\ee
since $\lag W \rag \simeq M_P^2 m_{3/2}$.
In addition, we can estimate the maximal order of magnitude of
$\lag G_{ai} \rag$
\be
  \lag G_{ai}\rag \simeq \lag - M_P^2 {W_a W_i \over W^2} \rag
           \simeq {f_{PQ}^2 \over M_P m_{3/2} }
\ee
since $\lag W_i\rag \simeq M_P m_{3/2}$ maximally and
$\lag W_a\rag \ler f_{PQ}^2$.  Therefore the condition
$\lag V_a\rag =0$ gives $\lag G_a\rag \ler f_{PQ}$.
In the axion direction $G_\Phi = \sum_a q_a v^a G_a /f_{PQ}$, we also have
$\lag G_\Phi\rag$ = $ \lag \sum_a q_a v^a K_a/ f_{PQ} \rag$ =
$\sum_a q_i |v^a|^2/f_{PQ} \ler f_{PQ}$.
Since the absence of accidental zero modes is required in the PQ--sector
one finds that $\lag G_a \rag \ler f_{PQ}$ for all the fields $S^a$.

It is now straightforward to estimate the maximal order of the axino mass
by using the fermion mass matrix in supergravity models, $\cm_{ij}$=
$M_P \exp(G/2M_P^2)[G_{ab} + G_a G_b/M_P]$.  Along the axino direction,
\be
  \cm_{\Phi b} \simeq m_{3/2} \left\lag \sum_a {q_a v^a \over f_{PQ}}
   G_{ab} - \sum_a {q_av^aG_a \over f_{PQ}} {G_b \over M_P} \right\rag\,.
\ee
{}From the $U(1)$ invariance of $G$, we get
$ \sum_a q_ava \lag G_{ab}\rag = q_b(v_b - \lag G_b\rag) \ler f_{PQ}$.
Hence the axino mass is maximally of order $m_{3/2}$.

The above order--of--magnitude estimation is indeed insensitive to the
specific forms of the kinetic term or the superpotential as long as the
hidden sector fields have only non--renormalizable couplings to the
observable sector. This happens because higher power terms in $K$
or $W$ are naturally suppressed by powers of $M_P$ which renders their
contribution negligible.  Therefore we conclude that the axino mass
in general supergravity models is at most of the order $m_{3/2}$ if
no other zero mode than the axion is present in the PQ--sector.

On the other hand if there are extra zero modes we are not able to
constrain further the order of $\lag G_a\rag$ for a zero mode direction $a$
so that the above argumentation breaks down. In fact, axino masses
$\gg\gr$ are possible in those models as we will see below.

\bigskip

We will now calculate the actual value of the axino mass relying on the
effective supergravity Lagrangian with arbitrary soft terms. This allows
to consider axino masses down to $O(\gr^2/f_{PQ})$ where the next to
leading order in the $1/M_P$--expansion of supergravity becomes important.

The superpotential $W$ of the PQ--sector is expanded as
\be
 W = f_a S^a + \frac{1}{2}f_{ab}S^a S^b + \frac{1}{6}f_{abc}S^a S^b S^c
\ee
and a departure from standard soft terms is encoded in
\be
 N = d_af_a s^a + \frac{1}{2}d_{ab}f_{ab}s^a s^b +
     \frac{1}{6}d_{abc}f_{abc}s^a s^b s^c \; .
\ee
Then the scalar potential reads
\be
 V = \db^b\bar{W}\p_b W + m_c^{2b}\bar{s}_bs^c+\gr\left[ s^b\p_b W
     +(A-3)W+N+{\rm h.c.}\right] \; .
\ee
To minimize this potential we apply the following strategy~: The VEVs $v^a$
are split into a global SUSY value $u^a$ with $\partial_a W(u)=0$ and
corrections $w^a$ due to the soft terms, $v^a=u^a+w^a$. Expanding
$\partial_a V$ around the global minimum $(u^a)$ results in
\bea
 (\p_aV)(v) &=& \frac{1}{2}f_{abe}\fb^{bcd}\wb_c\wb_dw^e
                +\frac{1}{2}M_{ab}\fb^{bcd}\wb_c\wb_d
                +\Mb^{bc}f_{abe}\wb_cw^e\nnu \\
            &&  +\frac{\gr}{2}(A+d_{abc})f_{abc}w^bw^c
                +\gr\left[((A-1)M_{ab}+d_{ab}f_{ab})w^b
              \right. \label{dV} \\
            &&\quad\quad\left. +(1+d_{abc})f_{abc}u^bw^c\right]+m^{2b}_a\wb_b
               +M_{ab}\Mb^{bc}\wb_c+I_a \nnu
\eea
with the global mass matrix
\be
M_{ab} = \p_a\p_bW(u)
\ee
and the definitions
\be
 I_a=\gr J_a+m^{2b}_a\bar{u}_b\; ,\quad J_a=M_{ab}u^b+(\p_aN)(u)\; .
 \label{J}
\ee
Care should be taken on the choice of $(u^a)$. With any global minimum
also a rotation $u^a\rightarrow\exp (q_az)u^a$ with $z=x+iy$ under
the complexified $U_{PQ}(1)$ leads to such a minimum. Despite the
$y$--dependent part of this symmetry which is clearly present in the
full theory the $x$--dependent part (present because of the
holomorphy of the superpotential) is broken by the soft terms. Therefore
an appropriate fixing for the $x$--dependent part of the symmetry should
be applied such that the global minimum comes close to the local values
$(v^a)$ resulting in small expansion coefficients $(w^a)$. Such a fixing
is provided by the condition $w_\a=\sum_aq_au^aw^a/f_{PQ} =0$ implying
that the correction in the global axion direction (denoted by an index $\a$)
vanishes.

As can be expected the axino mass is expressible in terms of the
corrections $(w^a)$~:
\bea
 (\cm\tilde{a} )_a &=& -\frac{1}{f_{PQ}}q_a^b(\p_bW)(v) \nnu \\
            &=& -\frac{1}{f_{PQ}}q_a^b\left( M_{bc}w^b+\frac{1}{2}
                           f_{bcd}w^cw^d+\cdots\right) \; .
\eea
These corrections have to be determined from eq.~\refs{dV}. Let us work in
a basis with diagonal global mass matrix $M_{ab}=M_a\d_{ab}$. We denote
massive modes with indices $i,j,\cdots$ and possible additional zero modes
with indices $\b ,\g ,\cdots$. Then an important observation is that the
corrections $w_i$ are basically determined by the linear term $|M_i|^2w^i$
and $I_i$ in eq.~\refs{dV}. In zero mode directions the situation might be
more complicated since $|M_\b |^2w^\b$ can be small compared to other terms
in eq.~\refs{dV}. Taking this into account we conclude that the expression
\be
 (\cm\tilde{a} )_a \simeq q_a^i \frac{\bar{J}_i}{f_{PQ}\bar{M}_i}\gr
                       -\frac{1}{f_{PQ}}q_a^\b M_\b w^\b
                       -\frac{1}{2f_{PQ}}q_a^\r f_{\r\b\g}w^\b w^\g
                       +O(\gr^2/f_{PQ}) \label{ax_mass}
\ee
gives the correct order of the axino mass. For the moment we leave the
values of $w_\b$ unspecified. Instead we concentrate on the first term
in eq.~\refs{ax_mass} and split the expression $J_a$ into its Planck
(or GUT)--scale value $J_a^{(0)}$ and corrections $J_a^{(1)}$ arising
{}from renormalization down to $f_{PQ}$.
The generic value of the corrections can be roughly estimated as
\bea
 J_a^{(1)} = \l_a 'k_a f_{PQ}^2 \label{J_rad}\\
 k_a \sim \frac{\l_a^2}{32\pi^2}\ln\left(\frac{M_P}{f_{PQ}}\right)
   \label{k_def}
\eea
with appropriate combinations $\l_a$, $\l_a '$ of the superpotential
couplings. According to eq.~\refs{J} $J_a^{(0)}$ is naively of the order
$f_{PQ}^2$. {\em Therefore a necessary condition for the axino mass to be
much smaller than $\gr$ is that $J_a^{(0)}=0$.}

In general this condition implies a relation between the structure of the
superpotential and the soft terms. It is instructive to analyze this
relation for a certain subclass of models, namely those with independent
soft coupling $A,B,C$ for the trilinear, bilinear and linear terms in
the superpotential, respectively. A computation leads to
\be
 J_a^{(0)} = (B-C)M_{ab}u^b+\frac{1}{2}(A-2B+C)f_{abc}u^bu^c\; .
\ee
Depending on the properties of the superpotential (and assuming that at
least one coupling $f_a$ is nonzero to force the symmetry breaking) several
cases can be distinguished~:
\begin{itemize}
\item $f_{ab}u^b$, $f_{abc}u^bu^c$, $M_{ab}u^b\ne 0$~: Then $J^{(0)}_a=0$
      if and only
      if $A=B=C$. No special property of the superpotential is required.
\item $f_{ab}u^b=0$, $Mu\ne 0$~: Then $J^{(0)}_a=0$ if and only if $A=C$.
      A simple superpotential which fulfills this requirement is
      e.~g.~$W=\l (SS' -\mu^2)Y$ since all $f_{ab}=0$.
\item $f_{abc}u^bu^c=0$, $Mu\ne 0$~: Then $J^{(0)}_a=0$ if and only if
      $B=C$.
\item $Mu=0$, $f_{ab}u^b\ne 0$~: Then $J^{(0)}_a=0$ if and only if
      $A-2B+C=0$.
      As the only ones these models allow for the full standard pattern
      $B=A-1$, $C=A-2$. They possess, however, at least one additional
      zero mode $\sum_au^aS^a$. An example is provided by the
      superpotential $W=\l (SS' -Z^2)Y-\l '(Z-\mu )^3$~\cite{ckn}.
\end{itemize}
Observe that in particular for $A=B=C$ the expression $J_a^{(0)}$
vanishes in any model.

If no additional zero mode is present the axino mass is already completely
determined by the first term in eq.~\refs{ax_mass}. On tree level this
means
\be
 \begin{array}{lll}
   J_a^{(0)} = 0 & \leftrightarrow & m^{(0)}_{\tilde{a}}=O(\gr^2/f_{PQ}) \\
   J_a^{(0)} \ne 0 & \leftrightarrow &  m^{(0)}_{\tilde{a}}=O(\gr )
 \end{array} \; .
 \label{crit}
\ee
We have therefore found a simple criterion to decide about the
magnitude of the axino mass which for soft terms specified by the
couplings $A,B,C$ singles out the particularly simple patterns listed above.
We remark that the small axino mass in no--scale models observed in
ref.~\cite{gy} can also be understood in terms of our analysis since in
those models $A=B=C=0$. A full supergravity computation of the axino mass
in no--scale models shows that their tree--level mass even vanishes.
Therefore the first line of eq.~\refs{crit} has to be understood as
a generic result which in certain special cases might be too large.
In the first case of a light axino mass radiative corrections to the
potential parameters can become important. Using eq.~\refs{J_rad} this
leads to a contributions of
\be
 m_{\tilde{a}}\simeq k_i\gr \; .
\ee
To keep the order $m_{\tilde{a}}=O(\gr^2/f_{PQ})$ an upper bound
on the couplings
has to be required. For $\gr\simeq 10^2$ GeV and $f_{PQ}\simeq 10^{11}$ GeV
this implies $\l_i\ler 10^{-4}$. A systematic way to avoid such small
couplings is to consider no--scale models.
Since gaugino masses are the only source of SUSY--breaking in those models
the standard model singlet fields in the PQ--sector will not receive any
radiative soft terms.

We see that without additional zero modes a complete answer can be
given. In particular we recover the result $m_{\tilde{a}}\ler O(\gr )$.

\bigskip

If additional zero modes are present the situation becomes more complicated
since the corrections $w_\b$ in the zero mode directions can become large.
In addition we have to consider that the zero entry $M_\b$ of the mass
matrix receives a radiative contribution $M_\b = \tilde{\l} '\tilde{k}_\b
f_{PQ}$ with $\tilde{k}_\b$ in analogy to eq.~\refs{k_def} unless it is
protected by an additional continuous or discrete symmetry.
Let us discuss some relevant cases. First we discuss a model
with $J^{(0)}_a=0$, e.~g.~a model of the last type in the above list for
$A,B,C$--type soft terms. Then a tree level axino mass $O(\gr^2/f_{PQ})$ is
not guaranteed as opposed to the case without additional zero modes~:
If the terms in eq.~\refs{dV} linear in $w_\b$ vanish (which e.~g.~occurs
for $f_{\b\g}=0$, $f_{\b\g\d}=0$ or $A=B$) a value $w_\b =O(I_\b^{1/3})$
results which causes an axino mass given by
$m_{\tilde{a}}^{(0)}=O((\gr/f_{PQ})^{1/3}\gr )$.
In any case a small axino mass $\ll\gr$ has to be stabilized against
radiative corrections. For superpotential couplings $\l =O(1)$ the axino
mass is shifted to $m_{\tilde{a}}=O(\gr )$.

Now we assume that $J^{(0)}_a\ne 0$. Then for values $\l =O(1)$ the
linear term $|M_\b |^2\wb_\b$ in eq.~\refs{dV} will dominate and the axino
mass is given by
\be
 m_{\tilde{a}}=O(\gr /\tilde{k}_\b )\; .
\ee
If, on the other hand, the couplings $\l$ are very small we can have
$w_\b =O(I_\b^{1/3})$ leading to an axino mass
$m_{\tilde{a}}=O((\gr^2 f_{PQ})^{1/3})$.
We see that axino masses $\gg\gr$ are indeed possible.
An example featuring  all these aspects is the
superpotential in the fourth entry of the above list for $A,B,C$--type
soft terms.

\bigskip

Now we turn to a discussion of the cosmological constraints on the masses
of the axino and the saxion.  We begin by noticing the fact
that self--couplings among the axion supermultiplet arise after integrating
out the heavy fields in the PQ--sector~:
\bea
 {\cal L} &=& \sum_i v_i^2 \exp[q_i(\Phi +\bar\Phi)/f_{PQ}] |_{\rm D-terms}
                             \nonumber\\
       &\sim & (1+{\sqrt{2} x \over f_{PQ}} s) \left( {1\over2}\partial_\mu
         a \partial_\mu a + {1\over2}\partial_\mu s \partial_\mu s
           + i\bar{\tilde{a}}\gamma_\mu \partial_\mu \tilde{a} \right)
   - {x\over f_{PQ}} \partial_\mu a \bar{\tilde{a}} \gamma_\mu \tilde{a} +
                 \cdots \, ,
\eea
where $x = \sum_i q_i^3 v_i^2/f_{PQ}^2$ .
In some cases, in particular in a model with superpotential
$W=\l (SS'-\mu^2)Y$ and universal scalar soft masses, $x$ is zero at the
Planck scale and receives a contribution
$x\sim \l^2\ln (M_P/f_{PQ})/64\pi^2$ when the RG--improved potential
at the PQ--scale is considered~\cite{ckl}. Generically, however, $x$ is of
order $1$. In this case the self--coupling becomes important since a saxion
can decay into two  axions faster than e.g., into two gluons.
Decay--produced axions do not heat the universe. Therefore the cosmological
effect of saxion decay is different from what has been investigated
assuming vanishing $x$~\cite{kim1,lyth}. If $x$ is of order 1, a stronger
bound on the saxion mass can be expected since the decay--produced axions
are not thermalized but red--shifted away.
The standard nucleosynthesis constrains the energy density of the
universe due to this red--shifted axions to be less than what is contributed
by one species of neutrinos at the time of nucleosynthesis.
This gives the constraint $m_s Y_s g_{*D} < T_{D}$ where
$T_{D} = 0.55 g_{*D}^{-1/4} \sqrt{\Gamma M_P}$ is the decay temperature of
the saxion and $\Gamma = x^2 m_s^3/ 8 \pi f_{PQ}^2$ its decay rate.
The relativistic degrees of freedom at $T_D$ are counted by $g_{*D}$.
We get
\be \label{slb}
  m_s > 2.4\; \TeV \left( f_{PQ}/x \over  10^{11}\; \GeV \right)^2
                \left( g_{*D} \over 100 \right)^{5/2} \,.
\ee
This bound on the saxion mass which receives a contribution $O(\gr )$
{}from scalar soft masses might be difficult to fulfill for large values
of $f_{PQ}$.

\bigskip

Cosmological implications of axinos were first discussed in ref.~\cite{rtw}
assuming unbroken R--parity. The axino mass can be constrained in two ways.
First, the axino can be the lightest supersymmetric particle (LSP).
Then it should be lighter than a few keV in order not to overclose
the universe.  Otherwise, the axino decays into at least one LSP composed
out of the neutralinos in the MSSM.  In this case,
the decay--produced neutralinos tend to overdominate the evergy
density of the universe.  To avoid this the axino
should be heavy enough to decay before the neutralinos decouple.
Considering the axino decay into photino plus photon, it was obtained that
the axino mass should be bigger than a few TeV~\cite{rtw}.  Then, the axino
decay into top quark and scalar top can be allowed.
{}From this decay channel one finds a less restrictive bound
\be \label{lb}
  m_{\tilde a} > 90\;  \GeV \left(m_\chi^0  \over 40\; \GeV \right)^2
                   \left(f_{PQ}/X_t \over 10^{11}\; \GeV \right)^2
                   \left(174\; \GeV   \over m_t \right)^2
                   \left(g_{*D} \over 100 \right)^{1/2} \,.
\ee
Here $X_t$ is the PQ--charge of the top quark.
We see that a wide range of axino masses between $O($keV) and $O(10^2$ GeV)
is excluded.

Let us now analyze how the above constraints modify if an inflationary
expansion is taken into account. The decoupling temperature of axino or
saxion is around the range of the reheating temperature $T_R \sim 10^{10}$
GeV which is the maximally allowed value to cure the gravitino problem
in supergravity models~\cite{ekn}. If decoupling of the axino occurs
before inflation the primordial axino relics are diluted away.
The above consideration, then,  has to be applied to the regenerated
population of axinos. We recall that the axino decoupling is determined
by its interactions with gluinos, quarks and anti--quarks via
gluon exchange~\cite{rtw}.  The axino decouples at the temperature
\be
  T_D \sim 10^9\; \GeV \left( f_{PQ} \over 10^{11}\; \GeV \right)^2
                   \left(0.1   \over \alpha_c \right)^3 \,.
\ee
The regenerated number density per entropy is given by~\cite{ekn}
\be
  Y \sim 2 \times 10^{-5} \left( 10^{11}\; \GeV \over f_{PQ} \right)^2
                     \left( T_R \over 10^{10}\; \GeV \right) \,.
\ee
Depending on the range of the axino mass we can distinguish three cases as
follows.

First, for a stable axino, the constraint from overclosure
gives the following loose bound on the axino mass in terms of $T_R$~:
\be \label{ubi}
  m_{\tilde a} < 160\; \keV \left( f_{PQ} \over 10^{11}\; \GeV  \right)^2
                     \left( 10^{10}\; \GeV  \over T_R \right) \,.
\ee
Second, an axino with a mass satisfying the lower bound in
eq.~\refs{lb} is still allowed by cosmology.
Finally, for an unstable axino with mass between the estimations in
eq.~\refs{ubi} and in eq.~\refs{lb}, one gets a bound on the reheating
temperature by replacing  the axino mass in eq.~\refs{ubi}
by the mass of the usual LSP since the decay--products of the axino
contain at least one LSP:
\be
  T_R \ler 2 \times 10^5\;\GeV \left(f_{PQ} \over 10^{11}\; \GeV \right)^2
                    \left( 40\; \GeV \over m_{\chi^0} \right)\; .
\ee
This represents a quite stringent bound on $T_D$.

\bigskip

In this letter we have analyzed the axino mass in general supergravity
models and the cosmological constraints on such models.
We have distinguished models with and without additional zero modes in the
PQ--sector. For the latter we found the axino mass to be at most of
$O(\gr)$.  In the context of an effective approach encoding supersymmetry
breaking in soft terms we were able to derive a simple necessary criterion
for a small tree level axino mass $\ll\gr$ given in terms of superpotential
and soft term properties. For uniform trilinear, bilinear and linear soft
couplings $A,B,C$ the criterion is always fulfilled for $A=B=C$ whereas
for $A=B\ne C$ or $A\ne B=C$ additional properties of the superpotential
had to be required. If the global vacuum $(u)$ represents an additional
zero mode of the globally supersymmetric theory, i.~e.~$Mu=0$ with the
global mass matrix $M$, the relation $A-2B+C=0$ which admits the standard
pattern $B=A-1$, $C=A-2$ is sufficient for the criterion to hold.

We showed that in models without additional zero modes our criterion is
sufficient, i.~e.~it guarantees a tree level axino mass of at most
$m_{\tilde{a}}^{(0)}=O(\gr^2/f_{PQ})$. In the presence of other zero modes
it serves as a good indication for such a small mass but additional
conditions (like e.~g.~$A\ne B$ for models with an $A,B,C$--pattern)
are required to have $m_{\tilde{a}}^{(0)}=O(\gr^2/f_{PQ})$.

{}From the cosmological point of view axino masses $O(\gr^2/f_{PQ})=O($keV)
are very attractive. In this case the axino is the LSP and can contribute
a relevant part of the mass in the universe as (warm) dark matter. Though
models with such an axino mass can be constructed as we have seen they
correspond to very special points in the space spanned by the
superpotential and soft term parameters. Moreover, such small masses are not
stable under radiative corrections arising from renormalization effects
between $M_P$ and the PQ--scale $f_{PQ}$. Taking these effects into
account the axino mass will be generically given by
$m_{\tilde{a}}^{(1)}= O(k\gr )$ with $k=\l^2\ln (M_P/f_{PQ})/32\pi^2$ and
a typical superpotential coupling $\l$.
The one--loop contribution from the characteristic coupling of the
axino to the Higgs--fields or heavy quarks will be given by
$m_{\tilde{a},{\rm loop}}\simeq 3g^2Am_{3/2}/16\pi^2$ which is only
relevant in the KVSZ--implementation of the axion.
As the LSP the axino has to be lighter than a few keV which in turn puts
a severe limit on the couplings $\l$ (and $g$ in the KVSZ--case),
typically $\l\ler 10^{-4}$. If the
decoupling temperature of the axino is larger than the reheating
temperature of inflation the overclosure bound on the regenerated
axino population is weakened resulting in a somewhat weaker bound on $\l$,
typically $\l\ler 10^{-3}$. In any case we conclude that a cosmological
relevant LSP--axino -- though possible in principle -- is not very likely to
occur~: Special models are needed and in addition small couplings have
to be chosen in order to avoid a conflict with the overclosure bound.

At this point it should be mentioned that no--scale supergravity models
can provide a naturally light axino~\cite{gy}.
Those models are characterized by a special pattern of the soft terms~:
The only non--vanishing soft terms at the Planck scale are gaugino
masses and therefore $A=B=C=0$ at tree level. Applying the above statements
a small axino mass results in this case.
In fact, a full supergravity calculation shows that the mass
vanishes on tree level. Other soft terms for the gauge non--singlets can be
generated due to renormalization group effects below the Planck scale in
those models. Since the PQ--sector consists of singlets their soft terms
are not affected by renormalization effects and the axino (saxion) remains
massless~\cite{gy}. Non--vanishing masses can be however
generated via the one loop contribution $m_{\tilde{a},{\rm loop}}$. In
the DFSZ--implementation they are so small that cosmological
effects of the axino and the saxino are negligible. This is clearly
different in the KVSZ--case. However, the limit on $g$ necessary to keep
the axino mass below the overclosure bound will be somewhat weakened with
respect to the ordinary case since the trilinear coupling $A$ for
$g\Phi Q_1Q_2$ originates from radiative corrections.

\bigskip

As a generic situation we consider models which do not fulfill our
criterion for a small tree level axino mass and possess couplings $\l=O(1)$.
The axino mass in such models is given by $m_{\tilde{a}}=O(\gr )$
(no additional zero modes) or $m_{\tilde{a}}=O(\gr /k)$
(additional zero modes). First of all this mass has to be larger than the
mass of the ordinary LSP in the MSSM to allow for a decay of axinos.
Second, these decays have to occur before the LSP decouples which translates
into a bound on $m_{\tilde{a}}$ of typically $m_{\tilde{a}}\ger 100$ GeV for
$f_{PQ}=10^{11}$ GeV if the decay channel into top and stop is possible.
Otherwise an even stronger
bound $m_{\tilde{a}}\ger$ TeV is required. We see that these generic models
are significantly constrained by cosmological considerations, however, a
final decision depends on details of the model like the exact axino mass,
the sfermion masses, the PQ--scale etc. Therefore it might be interesting
to study models which put further constraints on these parameters like
e.~g.~supersymmetric unified models incorporating an axion.
{}For a PQ--scale in the upper half of the allowed range $f_{PQ}>10^{11}$
GeV the lower bounds on $m_{\tilde{a}}$ and the saxion mass $m_s$ (which
both increase quadratically with $f_{PQ}$) become very stringent and it
might be difficult to construct viable models. In this context, having
axino masses larger than $\gr$ in models with additional zero modes might
be an interesting option.

\bigskip

{\bf Acknowledgement.}
This work was supported by the EC under contract no.~SC1-CT91-0729.
We would like to thank K. Choi and D. Matalliotakis for stimulating
discussions in the earlier development of this work.

\end{document}